\documentstyle[fleqn,epsf,11pt]{article}

\def\bo{{\raise.15ex\hbox{\large$\Box$}}}

\def\dag{^{\dagger}{}}

\def\ordless{{\lower2mm\hbox{$\,\stackrel{\textstyle <}{\sim}\, $}}}
\def\ordmore{{\lower2mm\hbox{$\,\stackrel{\textstyle >}{\sim}\, $}}}

\newtoks\slashfraction
\slashfraction={.13}
\def\slash#1{\setbox0\hbox{$\, #1$}
\setbox0\hbox to \the\slashfraction\wd0{\hss \box0}/\box0}

\def\leftrightarrowfill{$\mathsurround=0pt \mathord\leftarrow \mkern-6mu
        \cleaders\hbox{$\mkern-2mu \mathord- \mkern-2mu$}\hfill
        \mkern-6mu \mathord\rightarrow$}
\def\overleftrightarrow#1{\vbox{\ialign{##\crcr
        \leftrightarrowfill\crcr\noalign{\kern-1pt\nointerlineskip}
        $\hfil\displaystyle{#1}\hfil$\crcr}}}
\def\startarray{\left( \begin{array}}
\def\finarray{\end{array} \right)}
\def\starteq{
\begin{eqnarray}}
\def\fineq{\end{eqnarray}
}

\catcode`@=11 
\def\underline#1{\relax\ifmmode\@@underline#1\else
$\@@underline{\hbox{#1}}$\relax\fi}
\catcode`@=12
 
\newskip\humongous \humongous=0pt plus 1000pt minus 1000pt

\newif\ifdtup
 
\def\textcite#1{Ref.~{\cite{#1}}}

\def\thefootnote{\fnsymbol{footnote}}
 
\def\author#1#2{{\bf #1} \\ {\em #2}\vspace{5mm}}

\def\bold#1{\setbox0=\hbox{$#1$}%
     \kern-.025em\copy0\kern-\wd0
     \kern.05em\copy0\kern-\wd0
     \kern-.025em\raise.0433em\box0 }

\def\title#1#2#3#4#5{\thispagestyle{empty}
        \begin{center} \vspace*{1cm} { \bf #3} \\[.5in] {#4{}}
        \end{center} \vfill \centerline{ ABSTRACT}
   {\nopagebreak \noindent\begin{quotation}\noindent {\small #5}
   \end{quotation}} \vfill {#2} \hfill\begin{tabular}{r} {#1} 
        \end{tabular}  \newpage
        \def\thefootnote{\arabic{footnote}}}
\def\prefer{\section*{}
    \list{[\arabic{enumi}]}{\usecounter{enumi}\settowidth\labelwidth{[000]}
      \leftmargin\labelwidth\advance\leftmargin\labelsep \rightmargin=0pt}
        \small \sfcode`\.=1000\relax}

\def\refer#1{\section*{\large \sc {#1}}
    \list{\arabic{enumi}.}{\usecounter{enumi}\settowidth\labelwidth{[000]}
      \leftmargin\labelwidth\advance\leftmargin\labelsep \rightmargin=0pt}
        \raggedright \small \sfcode`\.=1000\relax}

\def\ReFer#1#2{\section*{\large\sc#1}
    \list{[\arabic{enumi}]}{\usecounter{enumi}\settowidth\labelwidth{#2} 
      \leftmargin\labelwidth\advance\leftmargin\labelsep \rightmargin=0pt}
        \raggedright \small \sfcode`\.=1000\relax}

\def\REFER#1#2{\section*{\large\sc#1}
    \list{#2 {enumi}.}{\usecounter{enumi}\settowidth\labelwidth{[000]}
      \leftmargin\labelwidth\advance\leftmargin\labelsep \rightmargin=0pt}
        \raggedright \small \sfcode`\.=1000\relax}

\def\startbib{\vspace{1in}\begin{refer}{References}
\small\frenchspacing\nopagebreak}
\def\endbib{\end{refer} \normalsize \nonfrenchspacing}
\def\startfig{\newpage \centerline{{\sl Figure captions}} \begin{itemize}}

\def\endfig{\end{itemize}}

\newcommand{\be}{\begin{equation}}
\newcommand{\ee}{\end{equation}}
\newcommand{\bea}{\begin{eqnarray}}
\newcommand{\eea}{\end{eqnarray}}

\newcommand{\AmS}{{\protect\the\textfont2
  A\kern-.1667em\lower.5ex\hbox{M}\kern-.125emS}} 
\begin{document}

\title{November 2004} {PC061.1104}
{Power corrections in charmless $B$ decays  
\footnotetext{${\dag}$  
Talk given at the QCD Euroconference 2004, Montpellier 5-9 July 2004}}
{\author{T. N. Pham} {Centre de Physique Th\'eorique, \\
Centre National de la Recherche Scientifique, UMR 7644, \\  
Ecole Polytechnique, 91128 Palaiseau Cedex, France}} 
{Power corrections seem to play an important role in charmless $B$
decays as indicated by recent analysis using QCD Factorization.
In this talk, I would like to report on a recent work
on  power corrections in charmless $B$ decays. By
using the ratio of the branching fraction of $B^+ \rightarrow \pi^+ K^{\ast 0}$ to
that of $B^0 \rightarrow \pi^- \rho^+$, for which the theoretical uncertainties are
greatly reduced, it is shown in a transparent manner that  
power corrections in charmless $B$ decays are probably large and  that the 
$B^0 \rightarrow K^- \rho^+$ decay could be explained with
the annihilation term included. For ratios of direct CP asymmetries, 
QCD Factorisation  with the annihilation terms included would
predict the direct CP asymmetry of $B \rightarrow \pi^+ \pi^-$ to be 
about 3 times larger than that of
$B \rightarrow \pi^\pm K^\mp$, with opposite sign. In particular, 
the large measured value 
for $B \rightarrow \pi^\pm K^\mp $ CP asymmetry implies naturally a corresponding 
large   $B \rightarrow \pi^+ \pi^-$ CP asymmetry as observed by Belle.
Experimentally any significant deviation from this prediction 
would suggest either new physics 
or possibly  the importance of long-distance rescattering effects.}

\section{Introduction}
In  QCD Factorization (QCDF)\cite{QCDF}, the $O(1/m_{b})$
 power corrections in  penguin matrix
elements and other chirally enhanced corrections could make important
contributions to the penguin-dominated charmless $B$ decays as in 
 $B \rightarrow \pi K$ decays.
Other power corrections terms such as annihilation contributions 
may also be present in $\rm PP$ and $\rm PV$ decays as first noticed 
in the perturbative QCD method for charmless $B$  decays\cite{pQCD}
and indicated by
 recent  analysis of charmless two-body non-leptonic
$B$ decays\cite{Du,Aleksan,Cottingham}. 
In a recent work\cite{Zhu}, we have shown that in QCDF, it is possible 
to consider certain
ratios of the $B \rightarrow PV$ amplitudes which depend only on the Wilson
coefficients and the known hadronic parameters. The discrepancy
between  prediction and experiment for the ratio would be a clear 
evidence for 
annihilation or other non factorisable contributions. We find that  
annihilation topology  likely plays an 
indispensable role at least for penguin-dominated 
$\rm PV$ channels. Including the annihilation terms in QCDF, we find that
 the direct CP asymmetry of $B \rightarrow \pi^+ \pi^-$ to be 
about 3 times larger than that of $B \rightarrow K^\mp\pi^\pm$, with opposite sign,
in agreement with experiment.  

\section{QCD factorization for charmless $B$ decays}

 The effective Lagrangian for non-leptonic $B$ decays
can be obtained from  operator product expansion and 
renormalization group equation, in which short-distance effects
involving large virtual momenta of the loop corrections from the scale
$M_W$ down to $\mu ={\cal O}(m_b)$ are  integrated into the Wilson
coefficients. The amplitude for the decay $B \rightarrow M_1 M_2$ can be
expressed as
\be
 {\cal A}(B \rightarrow M_1 M_2)=
 \frac{G_F}{\sqrt{2}} \sum_{i=1}^6 \sum_{q=u,c}
\lambda_q C_i (\mu) \langle M_1 M_2 \vert O_i (\mu) \vert B \rangle
\label{BMM1}
\ee
 $\lambda_q$ is a CKM factor, $C_i (\mu)$ are the Wilson coefficients
perturbatively calculable from first principles and  $O_{i}$ are the
 tree and penguin operators given by(neglecting other operators):
\bea
&&O_{1}= (\bar{s}u)_{L}(\bar{u}b)_{L} \quad, \ O_{4}= \sum_{q}(\bar{s}q)_{L}
(\bar{q}b)_{L}\nonumber \\
&&O_{6}= -2\sum_{q}(\bar{s}_{L}q_{R})(\bar{q}_{R}b_{L})
\label{Oi}
\eea 
The   hadronic matrix elements~:
$\langle M_1 M_2 \vert O_i (\mu) \vert B \rangle$  contains
 the physics effects from the scale 
$\mu ={\cal O}(m_b)$ down to $\Lambda_{\rm QCD}$.
 In the heavy quark limit, QCD Factorisation~\cite{QCDF} allows 
the  decay mplitude
$\langle M_1 M_2 \vert O_i (\mu) \vert B \rangle$ to  be factorized 
into hard radiative corrections
and  non perturbative matrix elements which can be parametrized by the
semi-leptonic decays form factors 
and meson light-cone distribution amplitudes (LCDAs). 

 Power corrections in $1/m_{b}$ come from penguin
matrix elements, chirally enhanced corrections and annihilation 
contributions. For example, in the $B \rightarrow \pi K$ amplitude, the 
matrix element of  $O_{6}$ is of the order $O(1/m_{b})$ compared to 
the $(V-A)\times (V-A)$ $O_{1}$ and $O_{4}$ matrix elements, since 
 $<K|\bar{s}_{L}d_{R}|0>$
is proportional to $m_{K}^{2}/m_{s}\approx 2.5\,\rm GeV$ while 
$<K|\bar{s}_{L}d_{L}|0>$ is proportional to  $K$ momentum which
is $O(m_{b})$, thus  numerically, the matrix element of $O_{6}$ which
has a factor
\be
r_\chi^K = \frac{2 m_K^2}{m_b (m_s + m_d)} \approx O(1)
\ee
is comparable to that of $O_{4}$. For penguin-dominated, 
decays, the $O_{4}$ and  
 $O_{6}$ matrix element are of the same sign in PP channnel, while in 
PV channel they are of opposite sign. Thus in QCDF one expects a small 
 $B \rightarrow K\rho$ branching ratio relative to $B \rightarrow \pi K$. Because of 
cancellation between the $O_{4}$ and $O_{6}$ contributions, the 
 $B \rightarrow K\rho$ decay
 is more sensitive to other power corrections and  non factorisable 
contributions. Including the 
chirally-enhanced corrections
in terms of  two quantities $X_{\rm A,H}$ and 
a strong phase, the   $B \rightarrow M_1 M_2$ decay amplitudes in QCDF
can be thus be written as\cite{QCDF1,QCDF2}:
\be
  {\cal A}(B \rightarrow M_1 M_2)=
 \frac{G_F}{\sqrt{2}}\sum_{p=u,c}V_{pb}V^{*}_{ps}  
 \left( -\sum_{i=1}^6 a_i^p
   \langle M_1 M_2 \vert O_i \vert B \rangle_f +
 \sum_{j} f_B f_{M_1}f_{M_2} b_j \right ),
\label{BMM}
\ee

\section{Power corrections in $B \rightarrow \rm PV$ decays}

Consider the  ratio of $A(B^+\rightarrow \pi^+ K^{\ast 0})$ to 
$A(B^0 \rightarrow \rho^+ \pi^-)$. amplitudes. If the power
corrections were negligible, this ratio would be theoretically very
clean where the form factors cancel out, furthermore it is
almost independent ot the  CKM angle $\gamma$ and the strange-quark mass:
 \be \label{pik}
 \left \vert \frac{{\cal A}(B^+ \rightarrow \pi^+ K^{\ast 0})}{{\cal A}(B^0 \rightarrow
\rho^+ \pi^-)}\right \vert \simeq 
 \left \vert \frac{ V_{cb}V_{cs}}{V_{ub}V_{ud}} \right \vert
\frac{f_{K^\ast}}{f_\rho} \left \vert 
\frac{a_4^c (\pi K^\ast)+r_\chi^{K^\ast} a_6^c (\pi K^\ast) }{a_1^u}
\right \vert 
\label{rhopi}
\ee

 $\vert (a_4^c (\pi K^\ast)+r_\chi^{K^\ast} a_6^c (\pi
K^\ast) ) /a_1^u \vert$ should be about or less than 0.04 in QCDF.
($f_{K^\ast}/f_\rho \approx 1$). The ratio 
 $\vert V_{ub}/V_{cb}\vert $  is not very well determined
 experimentally, but a stringent lower limit can be obtained 
from the unitarity of the
CKM matrix . Since \cite{Pham,Buras} :
\begin{equation}\label{inequality}
\left \vert \frac{V_{ub}}{V_{cb}}\right \vert = \lambda \sin \beta 
\sqrt{1+\frac{\cos^2 \alpha}{\sin^2 \alpha}} \ge \lambda \sin \beta ~.
\end{equation} 
and from the current Babar and Belle measured values
$\sin 2\beta= 0.736 \pm 0.049 $ \cite{HFAG1}~, we have
\begin{equation}\label{inequality2}
 \left \vert \frac{V_{ub}}{V_{cb}} \right \vert \ge \lambda \sin \beta =
0.090 \pm 0.007 > 0.078 
\end{equation}
Eq.(\ref{rhopi}) implies the following inequality~:
\be\label{pikrhopi}
 0.53 > {\left \vert \frac{{\cal A}(B^+ \rightarrow \pi^+ K^{\ast
0})}{{\cal A}(B^0 \rightarrow \rho^+ \pi^-)}\right \vert }= 0.77 \pm 0.09~,
\ee
where the number on the rhs is from the measured branching ratios
\cite{HFAG,laget}, . The lhs
would be reduced further to $0.46 \pm 0.04$, if one neglects a small 
$\cos^2 \alpha$ term  in Eq.(\ref{inequality}).

\begin{figure}[htb]
\centering
\leavevmode
\epsfxsize=8cm
\epsffile{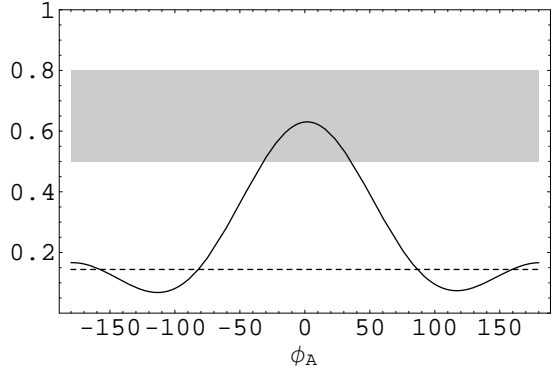}
\caption{ The ratio ${\cal B}(B^+ \rightarrow \pi^+ K^{\ast 0})/  
 {\cal B}(B^0 \rightarrow \rho^+ \pi^-)$ 
versus the weak annihilation phase
$\phi_A$. The default parameters are used but letting
the annihilation parameter $\rho_A=1$. The dashed lines show the
ratios without weak annihilation contributions. The gray areas denote
the experimental measurements with $1 \sigma$ error. }
\end{figure} 

\begin{figure}[htb]
\centering
\leavevmode
\epsfxsize=8cm
\epsffile{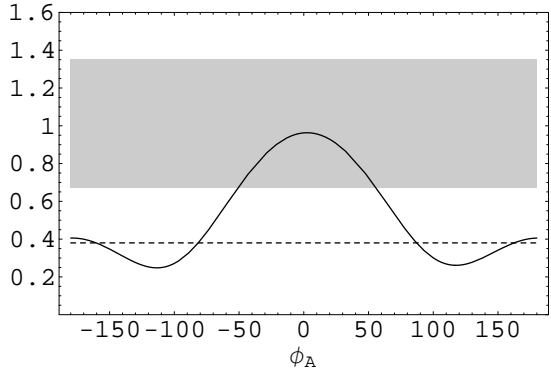}
\caption{ The ratio  
${\cal B}(B^0 \rightarrow K^+ \rho^-)/{\cal B}(B^0 \rightarrow \rho^- \pi^+)$
versus the weak annihilation phase
$\phi_A$. The default parameters are as in Fig.1 }
\end{figure} 

Since the  chirally enhanced corrections for penguin-dominated decays are
 not expected to be large, this large discrepancy 
 is   strong indication
that annihilation topology and/or other sources of power corrections might 
play an important role at least in $B \rightarrow \rm PV$ decays. There is  
 similar disagreement between theory and experiment in another ratio,
the branching fraction of $B^0 \rightarrow K^+ \rho^-$ to that of 
$B^0 \rightarrow \rho^- \pi^+$, though with large theoretical uncertainties.
For $\gamma=70^\circ$, $V_{ub}/V_{cb}=0.09$,
$a_4^c (\rho K)-r_\chi^K a_6^c(\rho K)=0.037+0.003 i$, $m_{s} = 90\,\rm MeV$,
we find 
\be
\frac{{\cal B}(B^0 \rightarrow K^+ \rho^-)}{{\cal B}(B^0 \rightarrow \rho^- \pi^+)} = 0.38
\ee
far below the measured value of  $1.01 \pm 0.34$, though, this 
ratio could be increased to $0.69$, if $m_{s}$ is lowered to 
$ 70\,\rm MeV$.

Taken together, these results indicate that
the penguin-dominated $B \rightarrow \rm PV$
decay amplitudes are consistently underestimated without annihilation 
contributions. Including the annihilation terms, from 
Eq. (\ref{BMM}), we have
\newpage
\bea
   A(B^+ \rightarrow \pi^+ K^{\ast 0}) &=& f_{K^\ast}F^{B \pi}m_B^2 a_4 +b_3(V,P)\nonumber \\
A(B^0 \rightarrow K^+ \rho^-) &=& f_K A^{B\rho}_0 m_B^2
(a_4-r_\chi^K a_6) + b_3(P,V)\kern 0.3cm
\eea
\be
  b_3 (M_1, M_2) = \frac{C_F}{N_c^2}\{ C_3 A_1^i (M_1,M_2)+   
 C_5 A^i_3 (M_1, M_2) +\kern -0.1cm (C_5+N_c C_6) A^f_3 (M_1, M_2) \}
\ee
With  the penguin
terms $a_4 \simeq -0.03$ and  $a_4-r_\chi^K a_6 \simeq 0.037$ having
opposite sign, the key observation is that $b_3 (V, P)$ and $b_3(P, V)$ ,
which get most of the contribution from $(C_5+N_c C_6) A^f_3$ term, 
are also roughly of the opposite sign since  $A^f_3(P,V)=-A^f_3(V,P)$.
Thus QCDF can easily enhance both ratios without fine
tuning ( no large strong phase ) as can be seen in Fig.2 .

\section{Direct CP violations} 
We now turn to the CP asymmetries in QCDF with annihilation terms included.
Because of the CKM factor and $SU(3)$ symmetry for the tree and
penguin matrix elements in   $B^{0} \rightarrow\pi^+ \pi^- $ and 
$B^{0}\rightarrow K^{+}\pi^-$ decays, one can derive a relation between 
direct CP asymmetries in these two channels. With the CP asymmetry given as:
\begin{eqnarray}
&&A_{\pi \pi}  =   
 \frac{4 \vert V_{ub}V_{ud}
V_{cb}V_{cd} T_{\pi\pi} P_{\pi\pi}\vert \sin \gamma \sin \delta }
{2 {\cal B}(B \rightarrow\pi^+ \pi^- )} ~\mbox{,} \nonumber \\
&&A_{\pi K} = -\frac{4 \vert V_{ub}V_{us} 
V_{cb}V_{cs} { T}_{\pi K} { P}_{\pi K}\vert \sin \gamma \sin {\tilde \delta} }
{2 {\cal B}(B \rightarrow\pi^+ K^- )}. 
\label{ACP}
\end{eqnarray} 
($\delta=\delta_P - \delta_T$ = strong phases difference between
the penguin and tree amplitudes), we find
\bea\label{DCPV}
 \frac{A_{\pi\pi}}{A_{\pi K}}&=&-\frac{f_\pi^2}{f_K^2}
\frac{{\cal B}(B \rightarrow\pi^+ K^- )}
{{\cal B}(B \rightarrow\pi^+ \pi^- )} \left \vert \frac{T_{\pi\pi}P_{\pi\pi}}
{{ T}_{\pi K} {P}_{\pi K}} \right \vert 
\frac{\sin \delta}{\sin {\tilde \delta}} \nonumber \\
& \simeq & (-2.7 \pm 0.3) 
\frac{\sin \delta}{\sin {\tilde \delta}} 
\eea
a consequence of the fact that  
$T_{\pi\pi}P_{\pi\pi}/{ T}_{\pi K} {P}_{\pi K}$ is close to $1$,
a reasonable approximation in QCDF, at about $10$ percent level uncertainty.
A previous derivation of this relation is given in \cite{Fleischer}.
Belle has claimed large direct CP asymmetry
observed in $B^0 \rightarrow \pi^+ \pi^-$ decay while BaBar has not confirmed
it yet, but both of them are close to a measurement on $A_{CP}(\pi^- K^+)$
\cite{Belle,BaBar}
\be
A_{\pi\pi}=\left \{ \begin{array}{ll}
0.58 \pm 0.15 \pm 0.07 & \mbox{(Belle)}~, \\
0.19 \pm 0.19 \pm 0.05 & \mbox{(BaBar)}~;
\end{array} \right. 
\ee

\be
A_{\pi K}=\left \{ \begin{array}{ll}
(-8.8 \pm 3.5 \pm 1.8)\% & \mbox{(Belle)}~, \\
(-13.3 \pm 3.0 \pm 0.9)\% & \mbox{(BaBar)}~.
\end{array} \right. 
\ee
We thus  expect  very 
naturally a larger direct CP violation
for $\pi^+ \pi^-$ decay compared with $\pi^- K^+$ decay, since 
the $\pi^+ \pi^-$ decay rate is smaller than
the $\pi^- K^+$ decay rate by  factor $3-4$, 

Experimentally, 
\begin{equation}
\frac{A_{\pi\pi}}{A_{\pi K}}=\frac{0.42 \pm 0.13}{-0.11 \pm 0.03}
  =-4.0 \pm 1.8 ~,
\end{equation}
still consistent with the theoretical estimation of $-2.7 \pm 0.3$.

Similar relation between CP asymmetries for the $B \rightarrow PV$ decays
for which the CP-violating interference terms are essentially of the same
magnitude, but with opposite sign:
\begin{eqnarray}
&&\frac{A_{\rm CP}(B^0 \rightarrow \rho^+ \pi^-)}{A_{\rm CP}(B^0 \rightarrow   K^{\ast +} \pi^-)}
\simeq  -\frac{{\cal B}(B^0\rightarrow   K^{\ast +} \pi^-)}{{\cal B}(B^0 \rightarrow 
\rho^+ \pi^-)}\frac{f_\rho^2}{f_{K^\ast}^2}\frac{\sin \delta_{\pi \rho}}
{\sin \delta_{\pi K^\ast}} \nonumber \\
&&\frac{A_{\rm CP}(B^0 \rightarrow  \rho^- \pi^+)}{A_{\rm CP}(B^0\rightarrow \rho^- K^+ )}
\simeq  -\frac{{\cal B}(B^0 \rightarrow 
  \rho^- K^+ )}{{\cal B}(B^0 \rightarrow  
\rho^- \pi^+)}\frac{f_\pi^2}{f_{K}^2}\frac{\sin \delta_{\rho \pi}}
{\sin \delta_{\rho K}}
\end{eqnarray}

\section{Conclusion}

 Power corrections in charmless B
decays are probably large, at least for the penguin-dominated PV channel.
 The key observation is that QCDF predicts 
the annihilation terms for $B^+\rightarrow \pi^+ K^{\ast 0}$ and $B^0 \rightarrow K^+ \rho^-$
to be   almost equal in  magnitude but opposite in sign and thus
enhance the decay rates for these two modes  to accommodate 
the experimental data. The relation for the direct CP asymmetry would 
naturally implies a large CP asymmetry
for $B \rightarrow \pi^+ \pi^-$ ,  about 3 times larger than that of
$B \rightarrow \pi^\pm K^\mp$ with opposite sign.

\bigskip

I would like to thank S. Narison and the organisers of QCD04 for the
warm hospitality extended to me at Montpellier.

\end{document}